\documentclass[fleqn,reprint,10pt,singlecolumn]{wlscirep}

\usepackage{setspace}
\usepackage{graphicx}
\usepackage{amsmath}
\usepackage{epstopdf}
\usepackage{letltxmacro}
\usepackage[english]{babel}
\DeclareMathOperator{\csch}{csch}
\LetLtxMacro{\ORIGselectlanguage}{\selectlanguage}
\makeatletter
\DeclareRobustCommand{\selectlanguage}[1]{%
  \@ifundefined{alias@\string#1}
    {\ORIGselectlanguage{#1}}
    {\begingroup\edef\x{\endgroup
       \noexpand\ORIGselectlanguage{\@nameuse{alias@#1}}}\x}%
}
\newcommand{\definelanguagealias}[2]{%
  \@namedef{alias@#1}{#2}%
}
\makeatother

\definelanguagealias{en}{english}
\definelanguagealias{English}{english}

\title{Thermal conductance of Nb thin films at sub-kelvin temperatures}

\author[1,*]{A. V. Feshchenko}
\author[1]{O.-P. Saira}
\author[1]{J. T. Peltonen}
\author[1]{J. P. Pekola}

\affil[1]{Low Temperature Laboratory, Department of Applied Physics, Aalto University, P. O. Box 13500, FI-00076 AALTO, Finland}

\affil[*]{anna.feshchenko@aalto.fi}

\begin{abstract}
We determine the thermal conductance of thin niobium (Nb) wires on a silica substrate in the temperature range of 0.1 - 0.6 K using electron thermometry based on normal metal-insulator-superconductor tunnel junctions. We find that at 0.6~K, the thermal conductance of Nb is two orders of magnitude lower than that of Al in the superconducting state, and two orders of magnitude below the Wiedemann-Franz conductance calculated with the normal state resistance of the wire. The measured thermal conductance exceeds the prediction of the Bardeen-Cooper-Schrieffer theory, and demonstrates a power law dependence on temperature as $T^{4.5}$, instead of an exponential one. At the same time, we monitor the temperature profile of the substrate along the Nb wire to observe possible overheating of the phonon bath. We show that Nb can be successfully used for thermal insulation in a nanoscale circuit while simultaneously providing an electrical connection.
\end{abstract}

\begin{document}

\flushbottom
\maketitle

\thispagestyle{empty}

\section*{Introduction}

Development of new technologies requires sustainable energy supplies \cite{Dresselhaus2007}. To this purpose, the field of thermoelectrics focuses on new materials and devices, such as heat engines \cite{Buttiker2013, Buttiker2014, Jordan2016} and refrigerators \cite{Pekola2007, Pekola2007a} for energy harvesting in low-dimensional systems \cite{Dresselhaus2007, Natthapon2010}. When it comes to mesoscopic devices fabricated at the nanoscale level, control of heat transport is essential \cite{Pobell2007} and challenging at the same time \cite{Pop2010}. These devices require sufficient and well defined thermal insulation between the hot and the cold parts to avoid unwanted heat exchange \cite{Pekola2007, Pekola2007a, Buttiker2014, Feshchenko2014}.

Superconductors are known to be good thermal insulators \cite{Pobell2007} and they are utilized for this purpose in a number of applications and research fields. 
Aluminium (Al) is one example of a thermal insulator that, however, works only in a limited range of temperatures up to about 300 mK \cite{Timofeev2009, Peltonen2010, Feshchenko2014}. Quasiparticle states that enable heat conduction begin to be populated at about 0.3 $T_{c}$, much before Al turns normal at the critical temperature of $T_{c}$ = 1.2 K. Often, however, one needs insulation in a wider range of temperatures. In this case, a possible candidate for a good thermal insulator is superconducting niobium (Nb), whose critical temperature is higher, 9.2~K in bulk. Nb has been used as a component  in a variety of mesoscopic devices, for instance, thermometers and coolers \cite{Maasilta2012}, quantum processors \cite{Orlando2007, Ustinov2007, Shcherbakova2015} and in space applications \cite{Stern2000}. In addition, superconducting resonators coupled to qubits are employed to detect and manipulate weak electromagnetic fields \cite{Schoelkopf2004, Schoelkopf2005, Tsai2007, Schoelkopf2008, Wallraff2008, Wallraff2011}. In a non-linear regime, resonators can be used for parametric amplification \cite{Haviland2007}. Ground and space-based commercial applications exploit Nb in photon detectors \cite{Klapwijk2007, Glowacka2008, Barends2009}. Incident radiation changes the temperature of a superconducting island in a transition edge sensor (TES) \cite{Giazotto2006, Karasik2008, Young2013} or its surface impedance in a kinetic inductance detector (KID) \cite{Mazin2004, Klapwijk2007, Glowacka2008}. Nb is a key component in many practical superconducting quantum interference devices (SQUID) \cite{Clarke2004}. As a particular example, in magnetoencephalography arrays of SQUIDS are used as very sensitive magnetometers for diagnostics of various diseases \cite{Montez2009}, and, in surgical planning for patients with brain tumors or epilepsy.

While so many devices are Nb based, we believe that in addition, Nb can be successfully used for careful engineering of the thermal environment to improve insulation or act as a heat switch in thermoelectric devices. Here, Nb is compatible \cite{Jabdaraghi2016} with the standard electron-beam lithography process used to fabricate tunnel junctions. This straightforward fabrication technique makes Nb attractive for multiple purposes. Furthermore, studying heat transport in Nb can help mitigate on-chip energy dissipation in quantum coherent devices \cite{Martinis2004}. Nb leads have been used for thermal insulation in circuits combining superconduting and normal state elements \cite{Roukes2005, Partanen2016}. Previous measurements of bulk thermal conductivity in superconducting single-crystal Nb have showed anomalous behavior that has been attributed to phonon conductance \cite{Carlson1970, Anderson1971}. However, no quantitative study of the residual electronic thermal conductance has been presented.

In this paper, we examine experimentally thermal conductance of Nb thin films (200 nm thick, 1 $\mu$m wide and 5, 10 and 20 $\mu$m long) in the superconducting state ($S$ state) in the temperature range of 0.1 - 0.6 K. We find that measured thermal conductance of Nb is 2-3 orders of magnitude lower than what is predicted in normal state ($N$ state) in this temperature range. It is approximately 2 orders of magnitude smaller than that in Al in $S$ state in the upper range of temperatures studied here (400 mK and above). However, we observe no exponential suppression of thermal conductivity even at these low temperatures. Instead, the thermal conductivity data is better described by a power law $T^{\beta}$, where $\beta \approx$ 4.5. 

Measurements of thermal conductance are performed using normal-metal-insulator-superconductor (NIS) tunnel junctions employed as heaters and thermometers. A similar technique has been used earlier in Ref. \citeonline{Peltonen2010} to study the thermal conductance of short Al wires influenced by the inverse proximity effect. In addition, we monitor the temperature of the phonon bath along the Nb strip by similar NIS sensors that are equally spaced on the substrate while being decoupled electrically from the main structure.

\section*{Theoretical background}

To elucidate the measurements of heat conduction along the Nb wire, we develop a basic thermal model. The heat flux through a conductor with constant cross section $A$ and length $L$ is given by
\begin{equation} 
\dot{Q} = \frac{A}{L}\int_{T_{1}}^{T_{2}}{\kappa(T)dT},
\label{Eq.1} 
\end{equation}
where $\kappa$ is thermal conductivity of the wire, and $T_{1}$ and $T_{2}$ are the temperatures at the two ends, respectively. Heat in metals is carried by electrons and phonons. However, at low temperatures the number of lattice vibrations, phonons, decreases and the diffusion of quasiparticles dominates the heat transport. For small temperature differences $T_{1} \approx T_{2}$, Eq.~(\ref{Eq.1}) can be linearized to 
\begin{equation} 
\dot{Q} = G(T) \Delta T,
\label{Eq.2}
\end{equation}
where $T = (T_{1} + T_{2})/2 $ is the mean temperature, $\Delta T = T_{2} - T_{1}$, and $G(T) = \frac{A}{L}\kappa(T)$ is the temperature dependent thermal conductance.

According to the Bardeen-Cooper-Schrieffer (BCS) theory of superconductivity, the electrons form a condensate of Cooper pairs that occupy the ground state below the material specific critical temperature, $T_{c}$. Quasiparticle excitations in the superconductor are separated from the ground state by the temperature-dependent energy gap, $\Delta$. The Cooper pairs cannot transport heat across the superconductor, unless they are depaired. Hence, the depaired electrons are the only source of heat transport and their number is exponentially suppressed at low temperatures $T \ll T_{c}$ making superconductor a good thermal insulator. Thermal conductivity in a typical superconductor can be written approximately as $\kappa_{S} = \kappa_{N}e^\frac{-\Delta}{k_{B}T}$ \cite{Timofeev2009}, where $\kappa_{N}$ is the normal state thermal conductivity, and $k_{B}$ is the Boltzmann constant. In the normal state, the thermal conductivity $\kappa_{N}$ = $L_{0}\sigma_{N}T$ depends linearly on temperature, and normal state electrical conductivity, $\sigma_{N}$, where $L_{0} = 2.44\times10^{-8}$ W$\Omega$K$^{-2}$ is the Lorenz number. In this experiment, we will demonstrate that thermal conductance of superconducting Nb nanostructures does not follow the predicted exponential dependence.
\begin{figure}[h!t]
\centering
\includegraphics[width=0.40\textwidth]{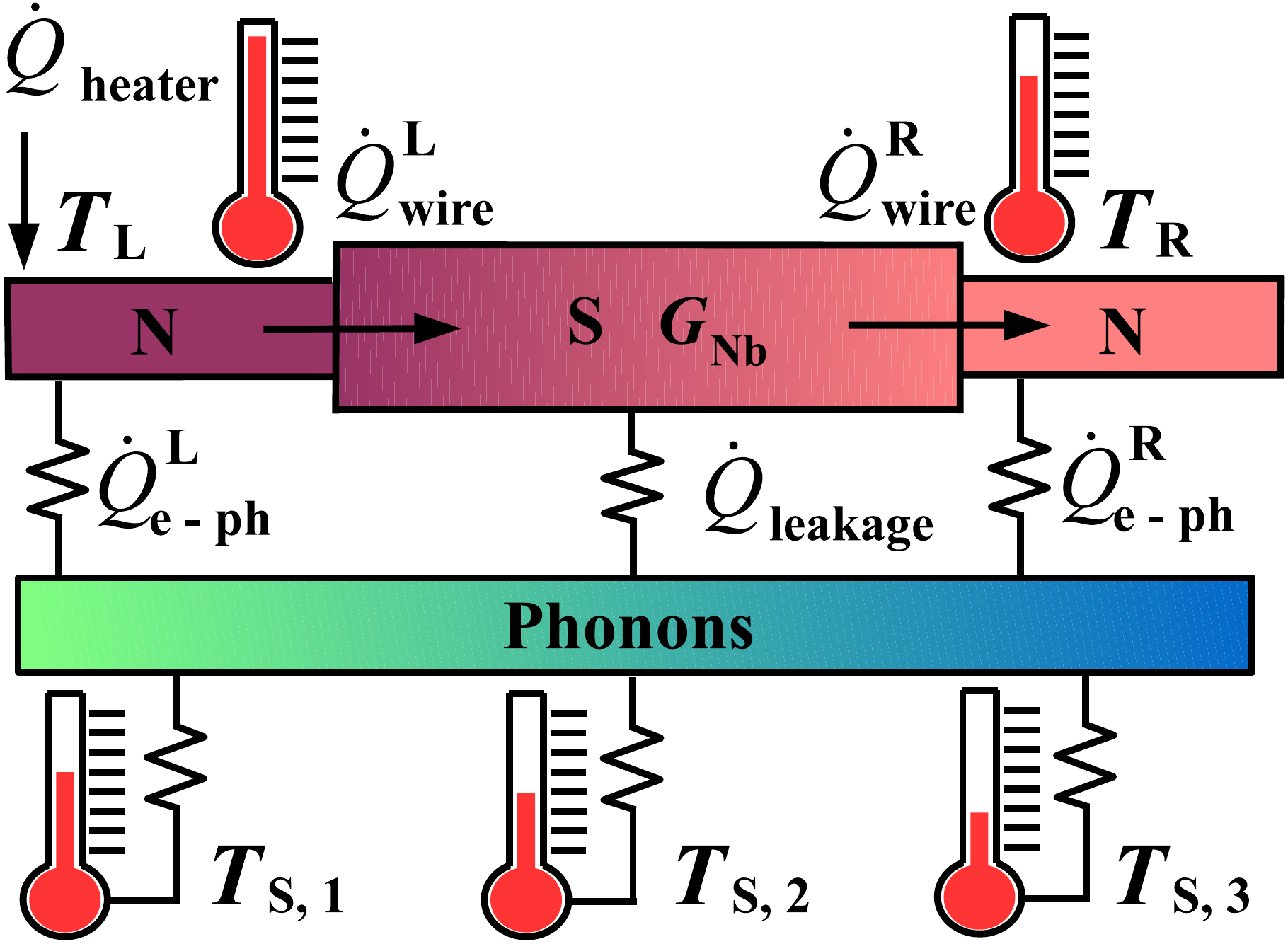}
\caption{Thermal scheme of the structure. Power applied to the heater is transmitted partially to the phonon bath via $\dot{Q}_{e-ph}^{L}$ and partially through the Nb strip to the right island, and then to phonon bath via $\dot{Q}_{e-ph}^{R}$. A fraction of the heat transported through Nb strip leaks to the substrate via $\dot{Q}_{leakage}$. We measure electronic temperatures at both ends of the Nb strip ($T_{L}$ and $T_{R}$). Temperatures $T_{S,1}$, $T_{S,2}$, $T_{S,3}$ of the phonon bath are monitored along the Nb strip. Nb strip is connected to the Cu islands at both ends via a clean ohmic contact.}
\label{fig:1}
\end{figure}
In Fig.~\ref{fig:1}, we show the thermal model we use to describe the heat flows in the present structure. The central part consists of two Cu islands that are connected with a Nb strip whose thermal conductance we study. In the experiment, we apply an electrical heating power $\dot{Q}_{heater}$ to the left ($L$) island and monitor the resulting temperature increase at various locations on the chip. The heater power flows from the left normal metal island to the phonon bath of the substrate via electron-phonon coupling $\dot{Q}_{e-ph}^{L}$ and to the Nb wire as $\dot{Q}_{wire}^{L}$. Part of the heat from the Nb strip leaks to the substrate through $\dot{Q}_{leakage}$ term and the rest is conducted to the right ($R$) normal metal island and eventually to the phonon bath, $\dot{Q}_{e-ph}^{R}$, via electron-phonon interaction. The temperature within the islands is assumed to be uniform due to low resistivity of Cu. At the same time, the phonon temperatures ($T_{S, 1}$, $T_{S, 2}$, and $T_{S, 3}$) are monitored by substrate thermometers alongside the Nb wire.

\section*{Results}

\begin{figure}[h!]
\centering
\includegraphics[width=\textwidth,height=7cm,keepaspectratio]{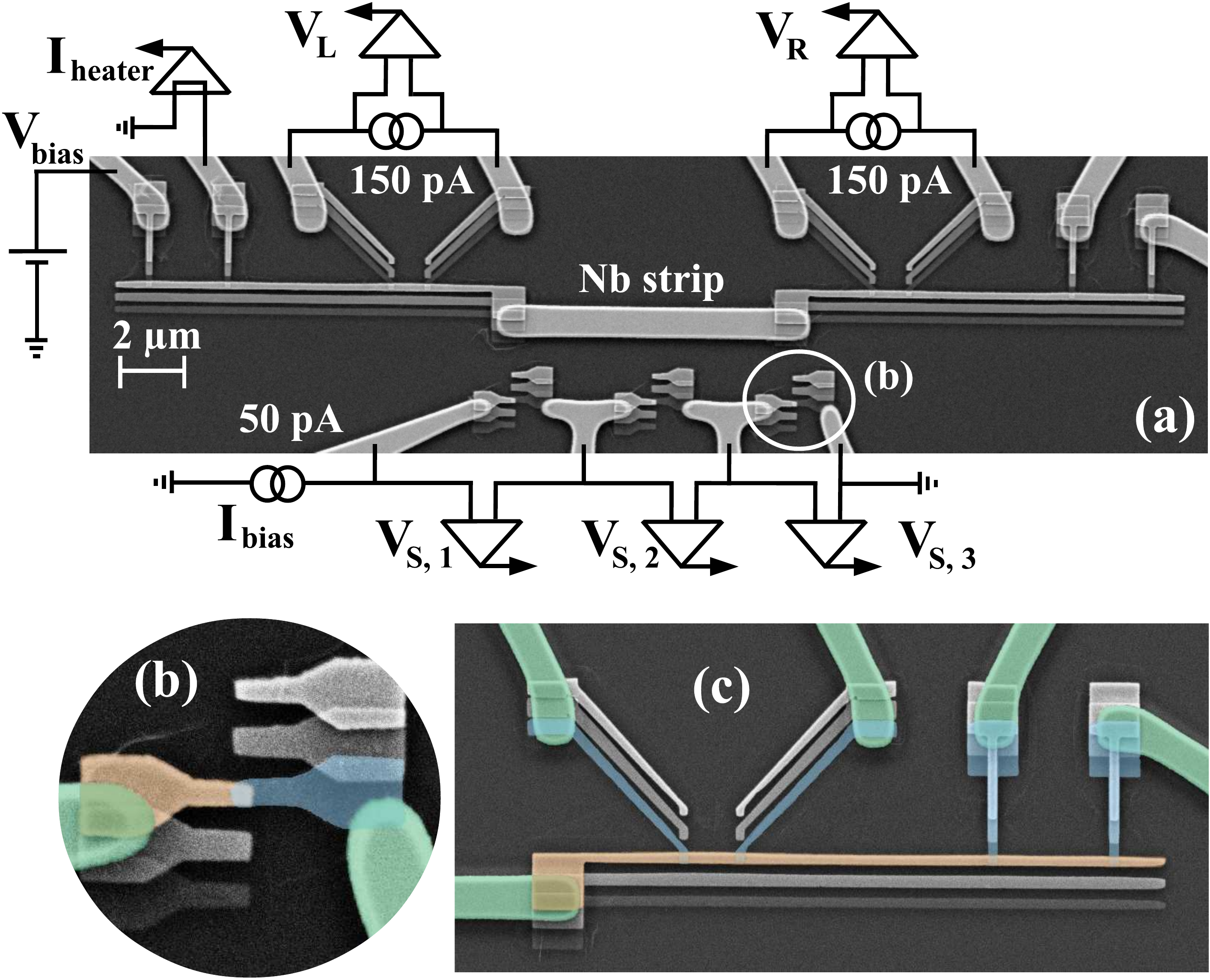}
\caption{(a) SEM image of the device $B$ shown together with a schematic of the experimental setup. The Nb strip is connected to two normal metal islands. The temperature of the left island is controlled by a pair of voltage-biased, $V_{bias}$, tunnel junctions. At the same time temperatures of the left and right normal metal islands are monitored by measuring the voltage drop ($V_{L}$ and $V_{R}$) across the two local thermometers. Similarly, the phonon bath temperature along the Nb strip is recorded with the help of the remote thermometers ($S 1$, $S 2$, and $S 3$). Close-up SEM image in (b) shows one of the remote thermometer junctions and (c) is an enlarged view of one local thermometer and heater. All Nb and Al leads are highlighted as light green and blue, respectively, while normal metal Cu islands are shown in beige.}
\label{fig:2}
\end{figure}

\begin{figure}[h!t]
\centering
\includegraphics[width=0.7\textwidth]{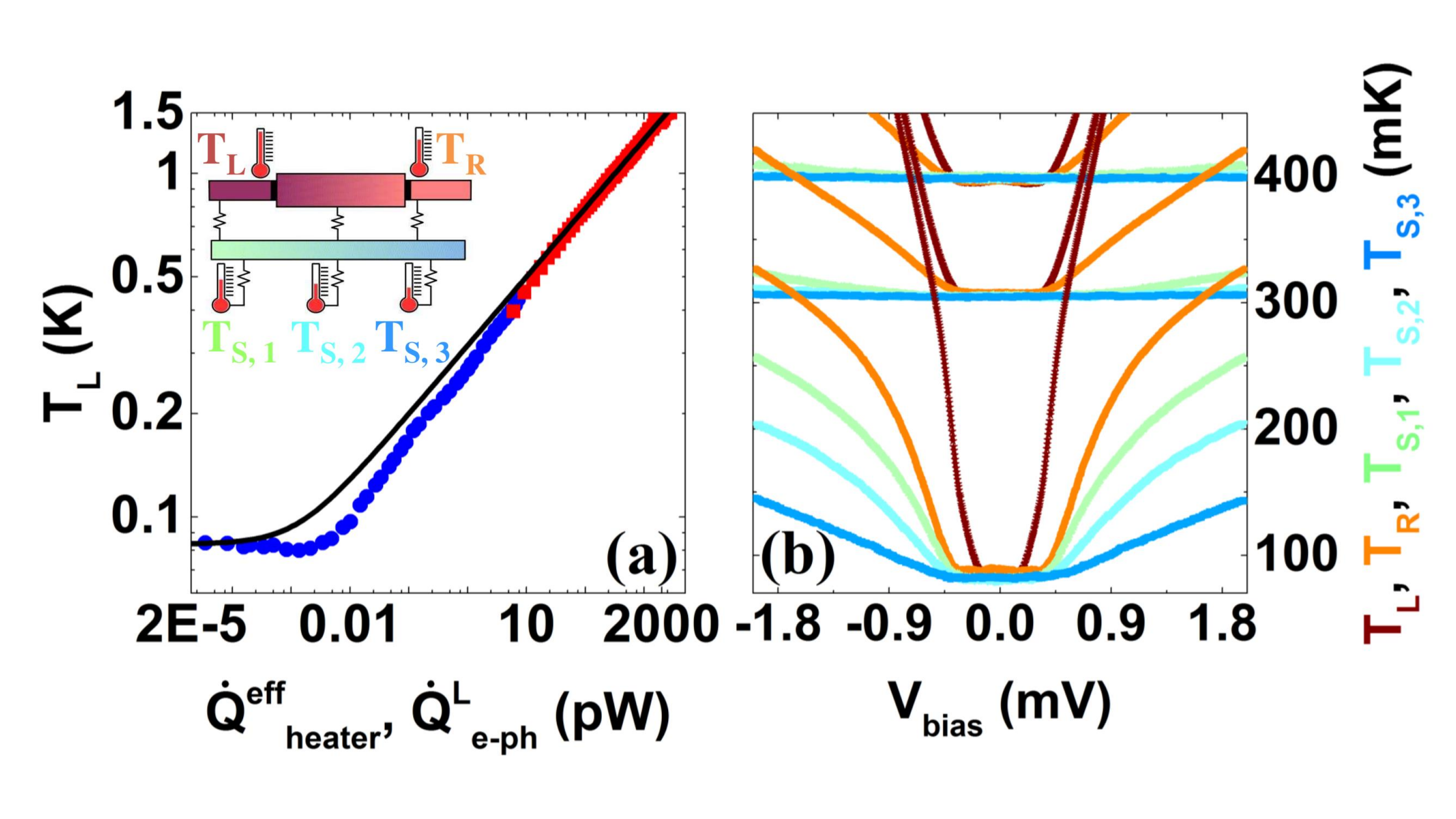}
\caption{In panel (a), two data sets (red squares and blue dots) correspond to two different types of measurements (see text) of sample $B$. The electronic temperatures, $T_{L}$, are extracted from a full fit of the $I-V$ curves (red points) or from the measured voltage across the current-biased thermometer using a polynomial calibration (blue points). The heater power for the red squares is calculated based on NIS tunnel junction theory presented in Eq.~\ref{Eq.5}, whereas the low heater power, $\dot{Q}^{eff}_{heater}$, for the blue dots needs special treatment (see Methods) due to NIS junction behaving as a cooler. The solid black line corresponds to the fit of electron-phonon power, $\dot{Q}_{e-ph}^{L}$, with a constant background heat load $\dot{Q}_{0}$, which is responsible for the saturation of electronic temperature. Measurements are performed at bath temperatures of $T_{bath}$ = 60 and 70 mK, resulting in an insignificant difference in the heat load. In panel (b), we show the temperature traces of the left ($T_{L}$, dark red), right ($T_{R}$, orange) and substrate ($T_{S, 1}$, green; $T_{S, 2}$, cyan; $T_{S, 3}$, blue) thermometers as a function of $V_{bias}$ measured at three different $T_{bath}$: 70, 305 and 400 mK.}
\label{fig:3}
\end{figure}

In this section, we describe the sample preparation, experimental techniques and analysis of the measured thermal conductance. We will refer to 5, 10 and 20 $\mu$m long Nb wire devices as $A$, $B$, and $C$, respectively. All the Nb wires are 1 $\mu$m wide and 200 nm thick. In the main panel of Fig.~\ref{fig:2}~(a), we show a scanning-electron micrograph (SEM) of the device $B$, and describe the main fabrication steps that are identical for all the devices. The device is fabricated in two lithography steps as in Ref. \citeonline{Jabdaraghi2016} allowing one to contact Nb with several other materials and employ shadow evaporation to realize tunnel junctions. First, 200 nm of Nb is sputtered on a SiO$_{2}$ substrate and after the resist spinning, all the Nb leads and Nb wire are defined by electron-beam lithography. Subsequently, after the development process, unnecessary metal is etched away by fluorine-based reactive-ion etching (RIE). The residual resist used to protect patterned Nb is removed by the following lift-off process. All the thermometers and heaters used to probe and control the temperature of the normal metal islands as well as the substrate thermometers are defined in the second lithography step. After this step, we use shadow angle evaporation technique to create metallization layers. For that, the native oxide of Nb is removed by $in-situ$ Ar plasma cleaning. Then the first 15 nm of Cu is evaporated at zero angle that forms a clean contact to Nb. Second layer of 20 nm of Al that is evaporated at an angle of 20 degrees creates all the thermometer and heater leads. After the oxidation of fresh Al $in-situ$ by thermal oxidation process, the last layer of 40 nm of Cu is evaporated at -20 degrees angle. This layer creates two long islands whose temperature is probed at both ends of the Nb wire, and normal state electrodes of the NIS tunnel junctions. A close-up of one of the substrate sensors is shown in Fig.~\ref{fig:2}~(b). In Fig.~\ref{fig:2}~(c), a close-up of two NIS junctions back-to-back (left) and a heater (right) is displayed. The false colors in Figs.~\ref{fig:2}~(b) and (c) reflect all Nb leads as green, Al leads as blue, and normal metal Cu island as beige.
The low-temperature measurements are performed in a $^3$He~-~$^4$He dilution refrigerator in the bath temperature range of $T_{bath}$ = 50$\dots$600 mK. In the present experiment, we perform two different types of measurements that are described below. In Fig.~\ref{fig:3}~(a), we study electron-phonon ($e-ph$) interaction between the left normal metal island and the substrate of sample $B$. The first data set (red squares) corresponds to a measurement where we control the temperature of the left island by the current through the heater $I_{heater}$, and record the full current-voltage ($I-V$) characteristic of the left thermometer for each value of the heating power. Here, the heater and the thermometer are NIS junctions whose current depends exponentially on the bias voltage and inverse temperature when the junction is biased slightly below the superconducting energy gap. Hence, one can calculate the power applied to the heater and extract the electronic temperature $T_{L}$ of the left island from fit to the $I-V$ characteristics without calibration against bath temperature as shown in Methods Fig.~\ref{fig:5}~(a). The second data set (blue dots) corresponds to a measurement where the heater is voltage biased at $V_{bias}$ and the left NIS thermometer is current-biased at 150 pA. The voltage drop ($V_{L}$) over the thermometer depends only on the electronic temperature of the island. Voltages measured across the NIS junction pairs are converted into electronic temperatures $T_{L}$ with a polynomial calibration \cite{Giazotto2006} with respect to the temperature of the dilution refrigerator ($T_{bath}$) as shown in Methods Fig.~\ref{fig:5}~(b). The two datasets are measured at $T_{bath}$ = 60 and 70 mK, respectively. The solid black line is a fitted $e-ph$ model $\dot{Q}_{e-ph}^{L}$ = $BT_{L}^{5}$ + $\dot{Q}_{0}$, where B and the background heating term $\dot{Q}_{0}$ = 0.74 fW are free parameters. Here, the saturation of electronic temperature $T_{L}$ comes from poor sensitivity of the NIS thermometer at low temperatures (see Methods Fig.~\ref{fig:5}~(b)), background heating $\dot{Q}_{0}$ and difficulty in modelling of $\dot{Q}_{heater}^{eff}$ precisely, however, this saturation is not important for measurements of thermal conductivity, as it is studied at temperatures above 0.1 K. The coefficient $B$ can be further factored as $B = \Sigma \Omega$, where $\Sigma$ is the material dependent electron-phonon coupling constant, and $\Omega$ is the volume of the metallic island. We find $\Sigma = 1.03\times10^{9}$ WK$^{-5}$m$^{-3}$ taking as the effective volume the two copies of the Cu island that result from the shadow evaporation technique. In the subsequent analysis where we evaluate the heat flow in the Nb wire, only the value of the coefficient $B$ is needed.
In Fig.~\ref{fig:3}~(b), we show the results of simultaneous temperature measurements of the left and right islands ($T_{L}$ and $T_{R}$) and substrate thermometers ($T_{S,1}$, $T_{S,2}$, and $T_{S,3}$). Again, we have converted the measured voltages into electronic temperatures with a polynomial calibration against the bath temperature as shown in Methods Fig.~\ref{fig:5}~(b). The heater of the right island is not used in this measurement. Temperature on the left island, $T_{L}$, increases more rapidly than on the right island, $T_{R}$, at any measured $T_{bath}$. At the same time, the substrate sensors ($T_{S, 1}$, $T_{S, 2}$ and $T_{S, 3}$) show elevated temperature of the phonon bath only at the lowest $T_{bath}$. The hierarchy of the magnitudes of temperature rise at the five positions follows the intuitive expectation based on the geometry and the connectivity of the sample. We quantify the temperature rise at the substrate locations in Methods Fig.~\ref{fig:6}. Substrate heating can be neglected at small power levels sufficient to study the wire heat conduction in linear response.

\begin{figure}[h!t]
\centering
\includegraphics[width=0.9\textwidth]{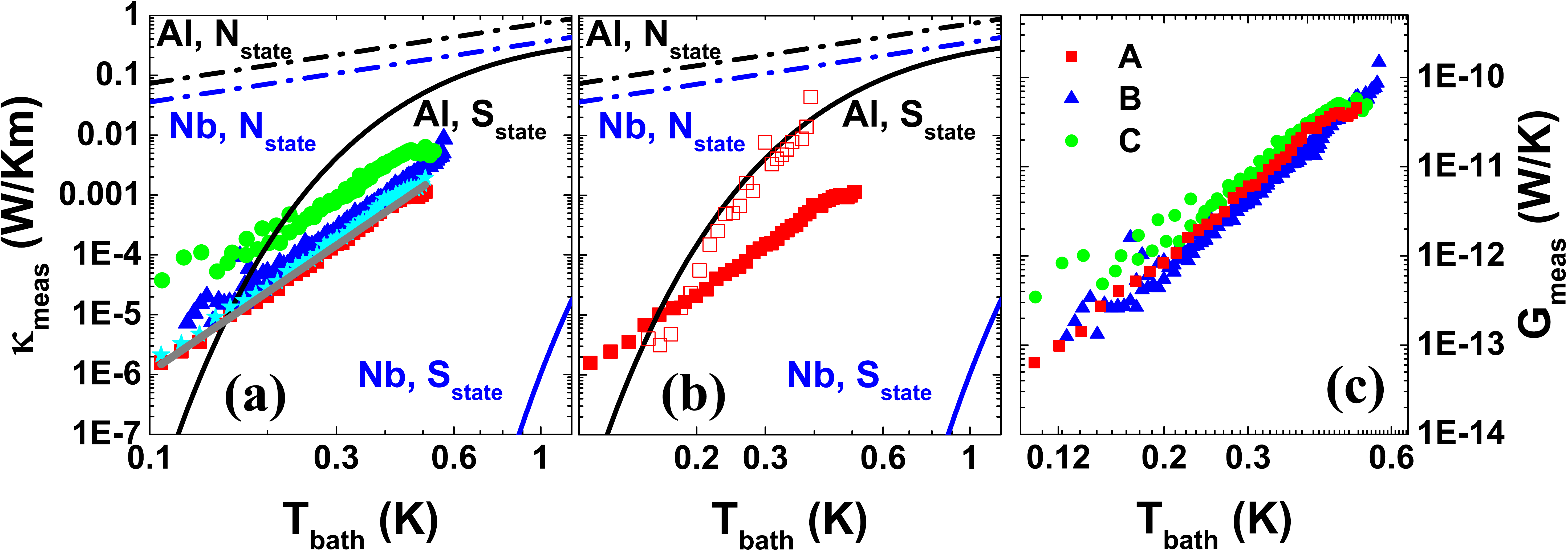}
\caption{In panel (a), the measured thermal conductivities of samples $A$, $B$ and $C$ are shown as red filled squares, blue triangles, and green circles, respectively. $\kappa_{meas}$ is obtained using thermal model as described in the main text. Bulk thermal conductivities of Nb and Al in the normal (dashed-dotted line) and superconducting (solid line) states are shown in blue and black, respectively. Solid grey line corresponds to a power law $T^{\beta}$, where $\beta \approx 4.5$. The cyan stars correspond to the extracted conductivity of sample $A$ using $IV$/2 as the injected power to the island. In panel (b), red open squares show a data set of thermal conductivity in Al wire from Ref. \citeonline{Peltonen2010} that is comparable in length to sample $A$ of Nb wire (red filled squares). In panel (c), we show samples $A$, $B$, $C$ as in (a), but in terms of the thermal conductances instead of conductivities.}
\label{fig:4}
\end{figure}

We can analyze quantitatively the observed temperature increases to extract the thermal conductance of the wire within the thermal model described above. In this measurement, we use a voltage bias, $V_{bias}$, for the heater and measure $I_{heater}$. For thermometers, we use constant current bias of 150 pA and record both $V_{L}$ and $V_{R}$. For the analysis, we write down two heat balance equations
\begin{equation}
\dot{Q}_{heater} = \dot{Q}_{e-ph}^{L} + \dot{Q}_{e-ph}^{R},
\label{Eq.3}
\end{equation}
where we omit the leakage term, $\dot{Q}_{leakage}$, for now and
\begin{equation}
\dot{Q}_{wire}^{R} = \dot{Q}_{e-ph}^{R}.
\label{Eq.4}
\end{equation}
The heater power is given by $\dot{Q}_{heater}$
\begin{equation}
\dot{Q}_{heater} = \frac{2}{e^{2}R_{T}}\int{n_{S}(E)(eV - E)\big[f_{N}(E-eV)-f_{S}(E)\big]}dE,
\label{Eq.5}
\end{equation}
where $R_{T}$ is the tunneling resistance of the junction, $E$ is the energy relative to the chemical potential, $n_{S}(E) = \Big|Re\big[\frac{E}{\sqrt{E^2 -\Delta^2}}\big]\Big|$ is a typical BCS density of states, and $f_{N,S}(E) = (e^{E/k_{B}T_{N,S}}+1)^{-1}$ is the Fermi distribution, where we assume $T_{N} = T_{S} = T_{L}$. A simpler model of the heater power where total dissipation at the heater junctions, $P_{total} = IV$, is split evenly between the superconducting and normal electrodes, i.e., $\dot{Q}_{heater}$ = $IV$/2, yields results that are almost indistinguishable from those based on this more sophisticated one (cyan stars in the Fig.~\ref{fig:4}~(a)). A full fit to the measured $I-V$ curves of the heater is used to extract a superconducting gap of the Al electrodes. For samples $A$, $B$ and $C$, we find superconducting gap and tunneling resistance for two heater junctions connected in series to be $\Delta$ = 214, 202 and 227.65 $\mu$eV, and $R_{T}$ =  56.5, 32 and 71 k$\Omega$, respectively.

At each bath temperature, we calibrate the electron-phonon flow by fitting the model $\dot{Q}_{heater} = B'(T_{L}^{5} + T_{R}^{5} -2 T_{bath}^{5})$ to the measurements. The value of $B'$ extracted in this manner is used to calculate the heat flow through the wire, $\dot{Q}_{wire}^{R}$, that is equal to $\dot{Q}_{e-ph}^{R}$ based on Eq.~\ref{Eq.4}. Here, the heat flow through the wire for small temperature differences is observed to follow an approximately linear dependence on the temperature difference across the wire, $T_{L} - T_{R}$. We now determine the thermal conductance $G_{meas}$ by a linear fit to the model $\dot{Q}_{wire}^{R}$ = $G_{meas}(T_{L} - T_{R})$ at each bath temperature using only data where the temperature difference falls between 3 and 15 mK. The data, fits, and more details on how the range is chosen are shown in Methods Fig.~\ref{fig:8}. Scaling $G_{meas}$ by the wire dimensions, we calculate the thermal conductivities that are shown in Fig.~\ref{fig:4}~(a) for sample $A$ (red squares), $B$ (blue triangles) and $C$ (green circles) in the range of $T_{bath}$ = 0.1 $\dots$ 0.6 K. For comparison, we show the expected thermal conductivity for bulk Nb and Al using the BCS expression for the $S$ state and Wiedemann-Franz Law for $N$ state. The experimentally observed thermal conductivity is not exponentially suppressed, but rather has a power law dependence $G_{meas} \propto T^{\beta}$, where $\beta \approx 4.5$. We show this power law as a solid grey line in the Fig.~\ref{fig:4}~(a). Figure~\ref{fig:4}~(c) shows the same data as thermal conductances instead. To show the comparison between Al and Nb thermal conductivity, we have added a reference data set analyzed based on the experiment from Ref. \citeonline{Peltonen2010} that is shown in Fig.~\ref{fig:4}~(b) as red open squares. Here, the Al wire is 4.2 $\mu$m long which is comparable with the Nb wire of 5 $\mu$m (sample $A$) that is marked as red filled squares. One can see that Nb has lower thermal conductivity at all temperatures above 0.2 K.

\section*{Discussion}

We now consider the effect of the heat leakage term, $\dot{Q}_{leakage}$, from the wire to the substrate on the previous analysis, where it has been ignored. First, the lack of $\dot{Q}_{leakage}$ term in Eq.~(\ref{Eq.3}) enlarges the calibrated $B$ and $B'$ prefactors for the electron-phonon coupling, as not all of the heating power flows to the substrate through the electron-phonon coupling in the islands. From all the measurements performed above, we find the prefactors $B$ or $B'$ for the electron-phonon coupling to be smaller than what is expected based on a standard value \cite{Giazotto2006} of $\Sigma$ for Cu and estimated volume of the structure. This indicates that the heat leak to the substrate from the wire, if present at all, does not dominate the $e-ph$ flow in the Cu islands. Moreover, as we show in Methods, the presence of the substrate leakage always reduces the measured thermal conductance of the wire. This conclusion holds also when the before mentioned effect on the calibration of the electron-phonon term is taken into account. Hence, the thermal conductance curves shown in Fig.~\ref{fig:4}~(c) are proper lower bounds, and our observation of anomalously high thermal conductance is not affected by the leakage.

In order to elucidate the influence of heat leakage from the Nb wire to the substrate on the apparent conductance, we study two simple models below. Here, we state the main results leaving the details of the calculations to the Methods. In the case of weak substrate leakage, a lumped element treatment is sufficient. We define the measured thermal conductance as $G_{meas}$ = $\frac{d\dot{Q}_{wire}^{R}}{dT_{L}}$, and find $G_{meas}$ = $\frac{1} {1+\rho /4}G_{Nb}$. Here, $\rho$ = $G_{S}/G_{Nb}$, and $G_{Nb}$ is the true thermal conductance through the Nb wire and $G_{S}$ is the thermal conductance to the substrate. In addition, the leak term can be calculated as $G_{leakage}$ = -$\frac{d\dot{Q}_{leakage}}{dT_{S}}$ = $\frac{1} {1+\rho /4}G_{S}$, assuming the substrate to be at a temperature $T_{S}$. At $G_{S} \simeq G_{Nb}$, the measured thermal conductance is still dominated by the heat conduction through the wire as $G_{meas}$ = $\frac{4}{5}G_{Nb}$. We may improve the above model by assuming a continuous temperature profile in the wire. Here, we find the measured thermal conductance to be $G_{meas} = \sqrt{\rho}\csch{(\sqrt{\rho})}G_{Nb}$. In the case $\rho \ll 1$, heat conductance through the wire dominates and $G_{meas} \approx G_{Nb}$. In the opposite limit of $\rho \gg 1$, one finds $G_{meas} \approx 2\sqrt{\rho}e^{-\sqrt{\rho}}G_{Nb}$.

Nevertheless, we do not know the precise dominant carrier of heat in Nb thin films, as it can be both electrons and phonons. In the present experiment these mechanisms are difficult to assess separately. Non-equilibrium quasiparticles studied in Al have been reported in a number of experiments \cite{Martinis2009, Saira2012, Riste2013, Pop2014, Taupin2016}. Similar heat carriers can be present in our experiment, however, the data is mainly taken at temperatures above 0.2 K where thermal quasiparticles are likely to play the dominant role.

In conclusion, we have measured thermal conductance of micron-scaled Nb wires over the temperature range of 0.1 - 0.6 K. We find that the thermal conductance exceeds the BCS predictions, and demonstrates a power law dependence on temperature as $T^{\beta}$, where $\beta \approx$ 4.5, instead of an exponential one. Still in the temperature range above 0.2 K Nb provides better thermal insulation as compared to e.g. aluminium.

\section*{Methods}

\subsection*{A. The extraction of electronic temperatures}

\begin{figure}[h!]
\centering
\includegraphics[width=0.70\textwidth]{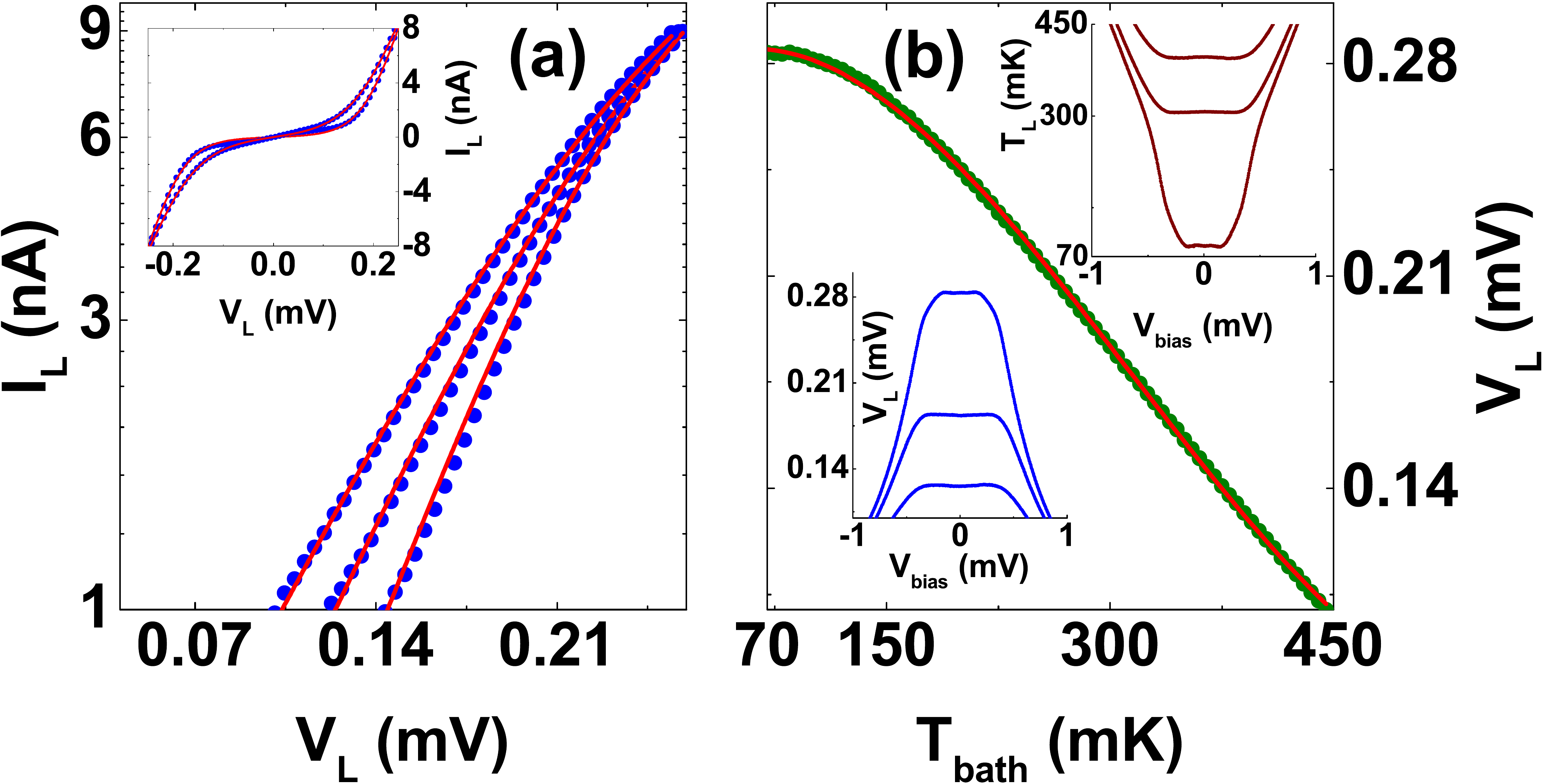}
\caption{In panel (a), we show three $I-V$ characteristics (blue dots) together with full fits (red lines) from which we extract $T_{L}$ = 591, 495 and 396 mK from left to right at $T_{bath}$ = 60 mK. In panel (b), we show the polynomial calibration with respect to $T_{bath}$ used to convert the voltage drop $V_{L}$ into electronic temperature $T_{L}$ (see text).}
\label{fig:5}
\end{figure}

The NIS tunnel junctions employed as the heater and thermometers have strong bias voltage and temperature dependence. The electronic temperature of the left island is assessed with two different methods as follows. In the first measurement where the temperature of the island is controlled by running the current through the heater, one can record the full $I-V$ characteristic of the left thermometer as shown in the inset of Fig.~\ref{fig:5}~(a). Here, solid red lines correspond to full fits to the measured $I-V$ characteristics utilizing
\begin{equation}
I(V) = \frac{1}{2eR_{T}}\int{n_{S}(E)\big[f_{N}(E-eV)-f_{N}(E+eV)\big]}dE,
\label{Eq.19}
\end{equation}
where $T_{L}$ and superconducting gap $\Delta$ are free parameters. In the main panel of Fig.~\ref{fig:5}~(a), we show several $I-V$'s (blue dots) on a logarithmic scale together with full fits (red lines) at $T_{bath}$ = 60 mK, where the three electronic temperatures extracted from the fits are $T_{L}$ = 591, 495 and 396 mK from left to right, and correspond to $T_{L}$ shown in Fig.~\ref{fig:3}~(a) as the red squares.
In the second measurement, the temperature of the island is controlled by $V_{bias}$ applied to the heater, while the left NIS thermometer is biased at a constant current of 150 pA, and its voltage drop $V_{L}$ is recorded simultaneously as shown in the lower inset of Fig.~\ref{fig:5}~(b). At zero bias voltage the temperature of the island is at $T_{bath}$, assuming no additional background heat load on the island. Thus with a polynomial calibration with respect to the temperature of the dilution refrigerator, shown in the main panel of Fig.~\ref{fig:5}~(b), one can convert measured voltages $V_{L}$ into electronic temperatures $T_{L}$. The upper inset of Fig.~\ref{fig:5}~(b) shows the dependence of the electronic temperature of the island on the bias voltage at three $T_{bath}$ = 400, 305 and 70 mK from top to bottom. These electronic temperatures $T_{L}$ correspond to those shown in Fig.~\ref{fig:3}~(b).

\subsection*{B. Effective power of the heater}

Throughout the manuscript, the heater power is calculated based on a basic result for an NIS tunnel junction presented in Eq.~\ref{Eq.5}. However, in Fig.~\ref{fig:3}~(a) for the data set marked as blue dots we calculate the effective heater power, $\dot{Q}^{eff}_{heater}$, using a different approach. This is due to the fact that the NIS junction behaves as a cooler at voltages close to the superconducting gap value. In addition, in the present structure it is difficult to model the cooling precisely due to complicated geometry of the islands and leads. Therefore, at low powers presented in Fig.~\ref{fig:3}~(a) as blue dots, we use the following empirical model \cite{Fisher1999, Koppinen2009, Chaudhuri2012}
\begin{equation} 
\dot{Q}^{eff}_{heater} = \dot{Q}_{heater} - \beta ' (\dot{Q}_{heater} - I_{heater}V_{bias}),
\label{Eq.20}
\end{equation}
where $\beta '$ is a dimensionless parameter and $0\leqslant \beta ' \leqslant 1$. Here, the first term is the effective cooling power of the normal metal island and the second term is the fraction of the heat flowing back to normal metal from the superconductor due to poor thermalization of the latter. We find $\beta '$ = 0.2 to be the optimum value to describe the heater behavior at low powers eliminating the cooling peak and yet at powers above 1 pW to match the basic NIS result of Eq.~\ref{Eq.5}. Furthermore, one can see that if $\beta '$ = 0, we get $\dot{Q}^{eff}_{heater} = \dot{Q}_{heater}$ that is again the result of Eq.~\ref{Eq.5}.

\subsection*{C. Measurement techniques: characterization of phonon temperature of the substrate}

\begin{figure}[h!t]
\centering
\includegraphics[width=0.40\textwidth]{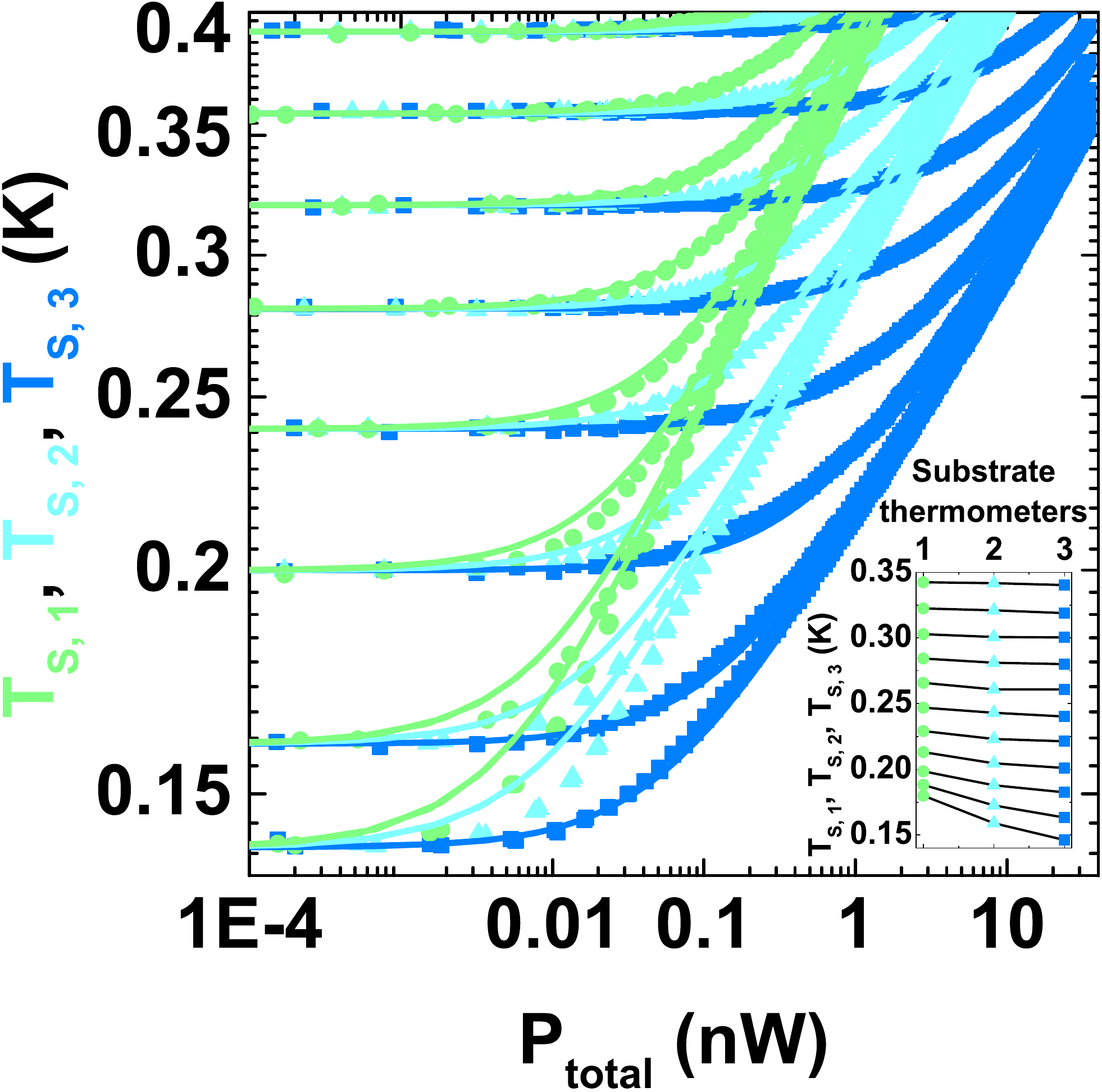}
\caption{Temperature of three substrate sensors versus the total dissipated power, $P_{total}$, at several $T_{bath}$. The measured temperatures $T_{S,1}$, $T_{S,2}$ and $T_{S,3}$ are plotted as green dots, cyan triangles, and blue squares, respectively. Corresponding solid lines are fits to the data (see text). The inset shows the temperature gradient along the substrate in the power range of 18-20 pW. Substrate sensors are positioned at the distances of $r_{S, 1}$ = 11.5 $\mu$m, $r_{S, 2}$ = 15.6 $\mu$m, and $r_{S, 3}$ = 20.4 $\mu$m from the heater, and show the strongest gradient at $T_{bath}$ $\leqslant$ 200 mK.}
\label{fig:6}
\end{figure}

In the main panel of Fig.~\ref{fig:6}, temperature traces versus the total dissipated power, $P_{total}$ = $I_{heater}V_{bias}$, for the three thermometers on the substrate are shown at several different bath temperatures. The solid lines correspond to a fit of $T_{S, fit} = \sqrt[n]{T_{bath}^{n} + P_{total}/b}$ to each set of curves, where $n$ and $b$ are the fit parameters \cite{Savin2006}. The substrate sensors are equally spaced along the Nb strip as shown in Fig.~\ref{fig:2}~(a). The temperature gradient along the Nb wire is observed by substrate sensors starting from $T_{bath}$ = 350 mK as illustrated in the inset of Fig.~\ref{fig:6}.

\subsection*{D. Characterization of Nb}

\begin{figure}[h!t]
\centering
\includegraphics[width=0.33\textwidth]{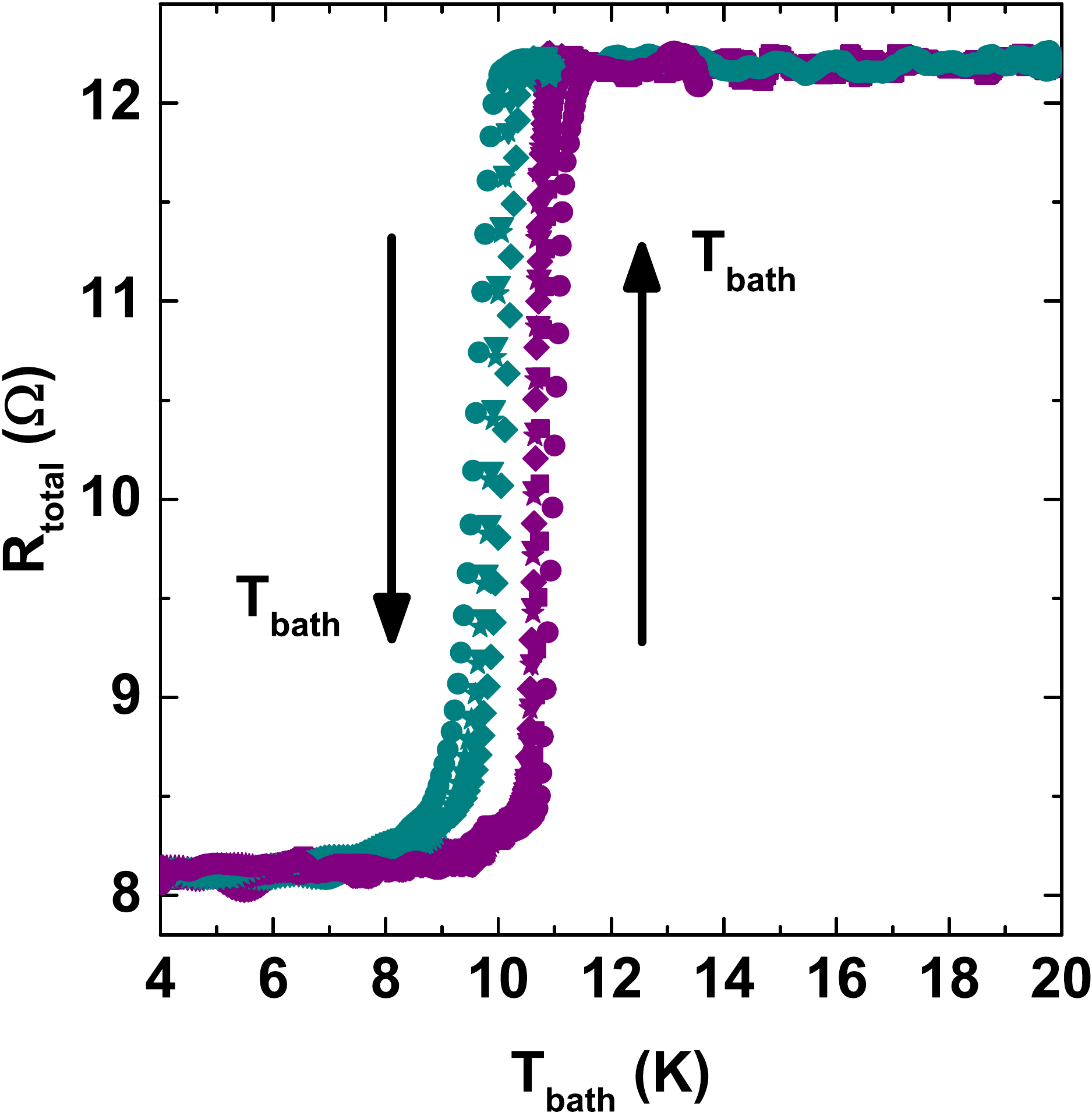}
\caption{The total series resistance $R_{total}$ versus T$_{bath}$ across the Nb strip. This resistance arises from a length of the copper island, two metallic contacts between the Cu and the Nb wire at its ends, and the resistance of the Nb wire itself in the normal state. The resistance of the Nb strip in the normal state is $R_{N}^{Nb} \approx 4~\Omega$, and can be estimated as the difference between the superconducting state ($R_{S} \approx 8~\Omega$) and the normal state of the series connections ($R_{N} \approx 12~\Omega$).}
\label{fig:7}
\end{figure}

The transition between the superconducting and the normal state was measured in the range of bath temperatures between 4 and 20 K. The measurement was done in a plastic dilution refrigerator insert immersed in liquid helium. To reach higher temperatures up to 20 K, the sample stage was electrically heated. Due to large temperature gradients in the dilution refrigerator, the measured resistance shows a hysteretic behavior. However, the transition occurs between 9 and 10 K, consistent with $T_{c}$ of bulk Nb as shown in Fig.~\ref{fig:7}.

\subsection*{E. Analysis of measured thermal conductance}

\begin{figure}[h!t]
\centering
\includegraphics[width=0.7\textwidth]{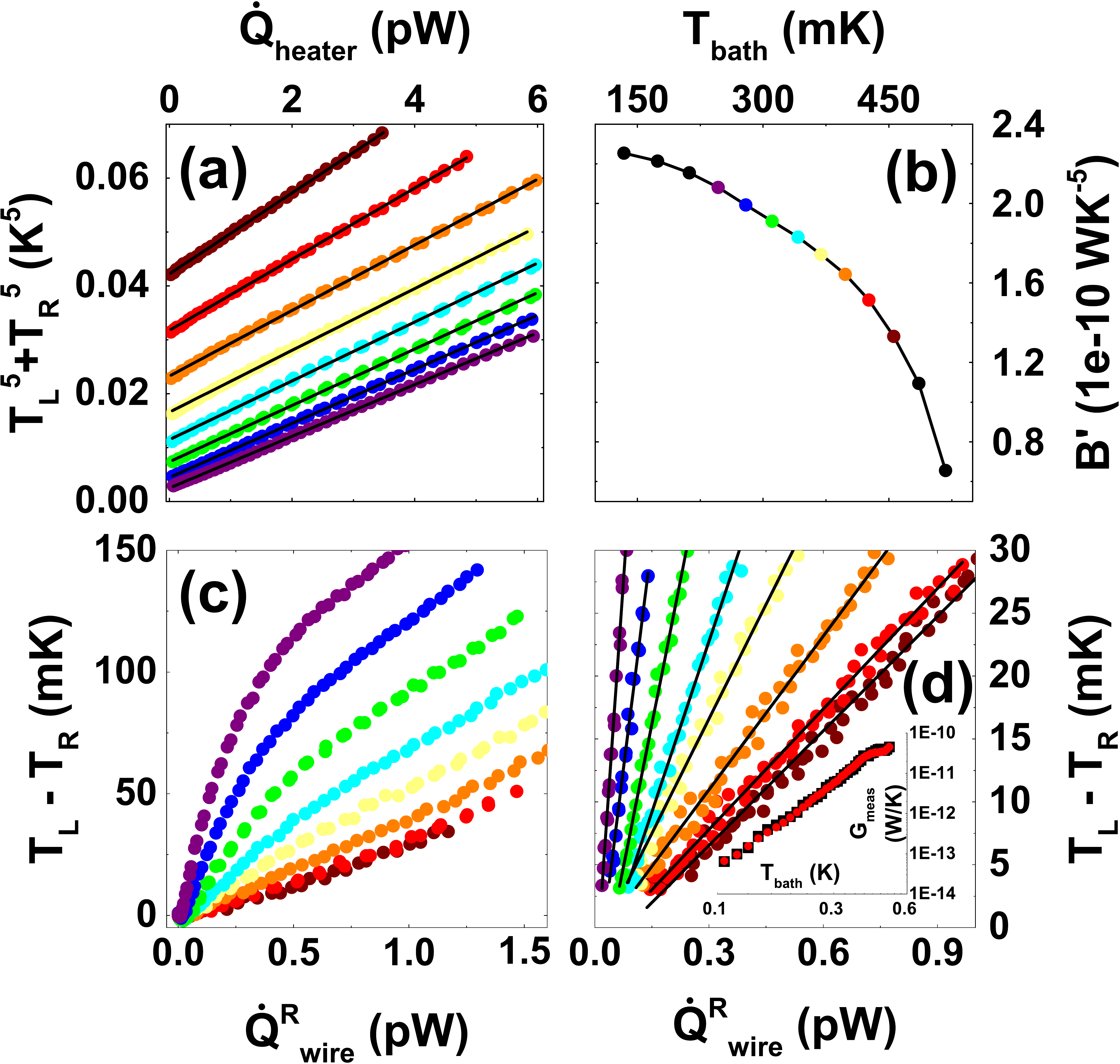}
\caption{In panel (a), we show the heater power as dots at several $T_{bath}$ together with linear fits (black solid lines) used to extract the electron-phonon coefficient B'. In panel (b), we show the extracted B' parameter at several $T_{bath}$. In panel (c), the heat flow through the wire versus the temperature difference is shown at several $T_{bath}$. In panel (d), a close-up into the linear regime of panel (c) is shown together with linear fits (solid black lines). The inset shows the extracted thermal conductance using an upper limit of 15 (black squares) and 30 (red circles) mK for the fitting range.}
\label{fig:8}
\end{figure}

Once we know precisely the heater power $\dot{Q}_{heater}$ via Eq.~(\ref{Eq.5}), and the island temperatures $T_{L}$, $T_{R}$ via calibration of the NIS thermometers as shown in Fig.~\ref{fig:5}~(b), we can estimate the heat flow through the Nb wire that reaches the right island using Eq.~(\ref{Eq.4}). First, we calibrate the electron-phonon coefficient B' utilizing a heat balance equation as $\dot{Q}_{heater} = B'(T_{L}^{5} + T_{R}^{5} -2 T_{bath}^{5})$. We show the data (points) together with fits (solid lines) at several $T_{bath}$ in Fig.~\ref{fig:8}~(a).
The extracted parameter B' versus $T_{bath}$ is shown in Fig.~\ref{fig:8}~(b). Once we know B', we can calculate the heat reaching right island via $\dot{Q}_{wire}^{R} = \dot{Q}_{e-ph}^{R}$, where $\dot{Q}_{e-ph}^{R} = B'(T_{R}^{5} - T_{bath}^{5})$. In Fig.~\ref{fig:8}~(c), we plot the heat flow through the wire, $\dot{Q}_{wire}^{R}$, versus the temperature difference $\Delta T = T_{L} - T_{R}$ at several $T_{bath}$. A close-up into the linear regime for small temperature differences is shown in Fig.~\ref{fig:8}~(d). We only exclude data where the temperature difference is smaller than 3 mK due to an anomalous behavior observed at temperature differences below this threshold (this is due to the fact that here the heater is biased at low voltages where its $I-V$ curve is highly non-linear). For the upper bound, we have used two values, 15 and 30 mK, to check the sensitivity of the fit. A linear fit to the data (solid black line) within these limits gives the measured thermal conductance, $G_{meas}$. The extracted thermal conductance curve using an upper limit of 15 and 30 mK is shown in the inset of Fig.~\ref{fig:8}~(d). Obtained conductances do not depend on the choice of the fit range.

\subsection*{F. Three-resistor thermal model}

\begin{figure}[h!t]
\centering
\includegraphics[width=0.35\textwidth]{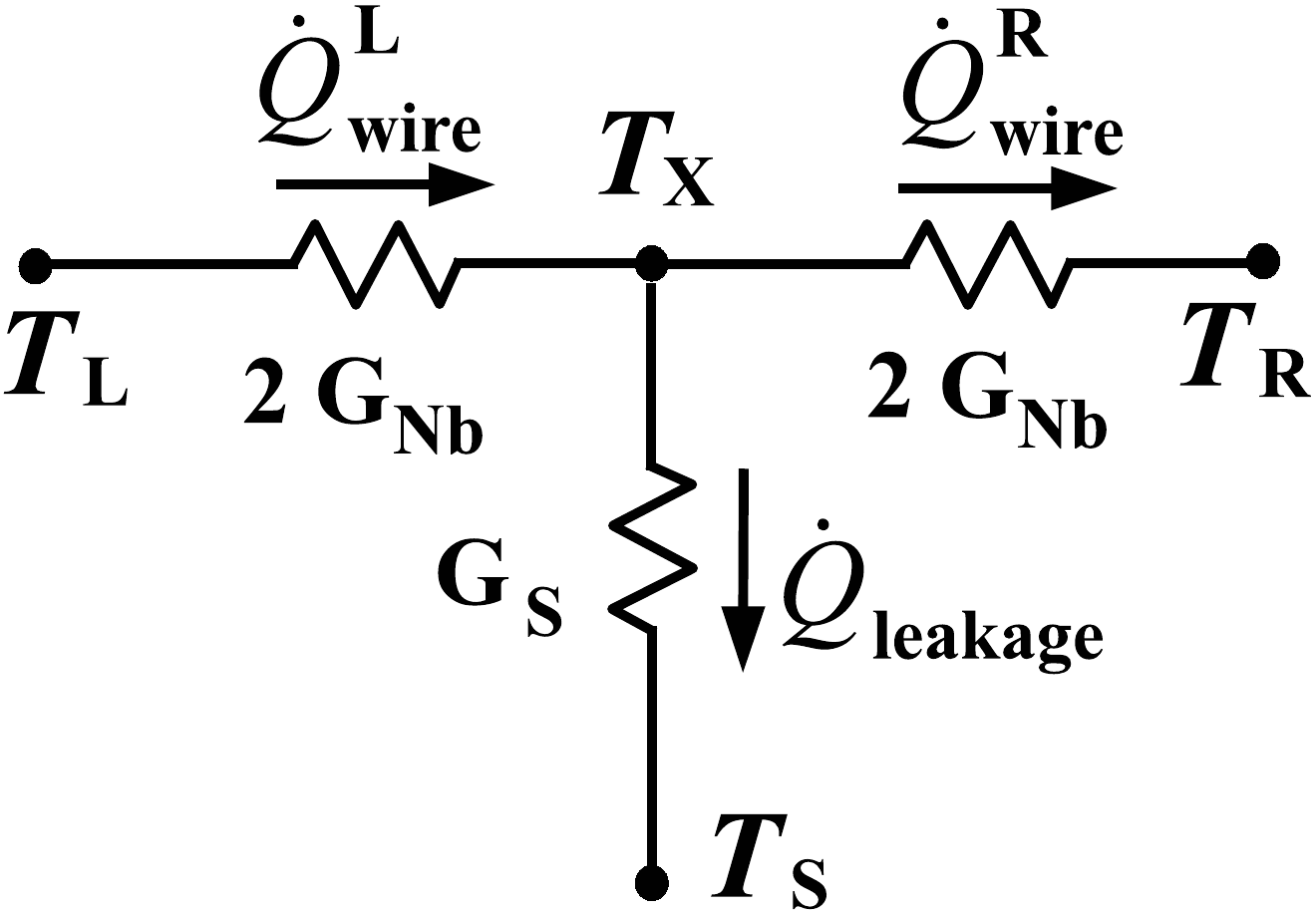}
\caption{Three-resistor model.}
\label{fig:9}
\end{figure}

For small temperature differences, and assuming the substrate leak is weak compared to the thermal conductance of the wire, the heat flows in the structure can be approximated by the lumped element three-resistor model shown in Fig.~\ref{fig:9}.
Here, we write down the Kirchhoff's law for the heat currents as
\begin{equation}
\begin{split}
& \dot{Q}_{wire}^{L} =  2G_{Nb}(T_{L} - T_{x}) & \\
& \dot{Q}_{wire}^{R} = 2G_{Nb}(T_{x} - T_{R}) & \\
& \dot{Q}_{leakage} = G_{S}(T_{x} - T_{S}) & \\
& \dot{Q}_{wire}^{L} = \dot{Q}_{leakage} + \dot{Q}_{wire}^{R}. &
\label{Eq.6}
\end{split}
\end{equation}
Solving the system of linear equations, we find the heat flow at the ends of the Nb wire and a leakage term as 
\begin{equation}
\begin{split}
& \dot{Q}_{wire}^{L} = \frac{2G_{Nb}}{4 + \rho}\big[(2 + \rho)T_{L} - 2T_{R} - \rho T_{S}\big] & \\
& \dot{Q}_{wire}^{R} = \frac{2G_{Nb}}{4 + \rho}\big[2T_{L} -  (2+\rho)T_{R} + \rho T_{S}\big] & \\
& \dot{Q}_{leakage} = \frac{2}{4 + \rho} \rho G_{Nb}\big[T_{L} + T_{R} - 2T_{S}\big]. & \\
\label{Eq.7}
\end{split}
\end{equation}
The measured quantity is the heat conducted through the wire that reaches the right Cu island. In the present experiment, we cannot independently determine the heat conductance through the Nb wire and heat leak to the substrate. Within the lumped-element model, the effective conductances are related to the conductances of the circuit as
\begin{equation}
G_{meas} = \frac{d\dot{Q}_{wire}^{R}}{dT_{L}} = \frac{4G_{Nb}}{4+\rho},
\label{Eq.8}
\end{equation}
and
\begin{equation}
G_{leakage} = -\frac{d\dot{Q}_{leakage}}{dT_{S}} = \frac{4 G_{S}}{4 + \rho}.
\label{Eq.9}
\end{equation}
For $\rho \approx 1$, one gets $G_{meas} \approx \frac{4}{5}G_{Nb}$ and $G_{leakage} \approx \frac{4}{5}G_{S}$.

\subsection*{H. Continuous thermal model}

\begin{figure}[h!t]
\centering
\includegraphics[width=0.35\textwidth]{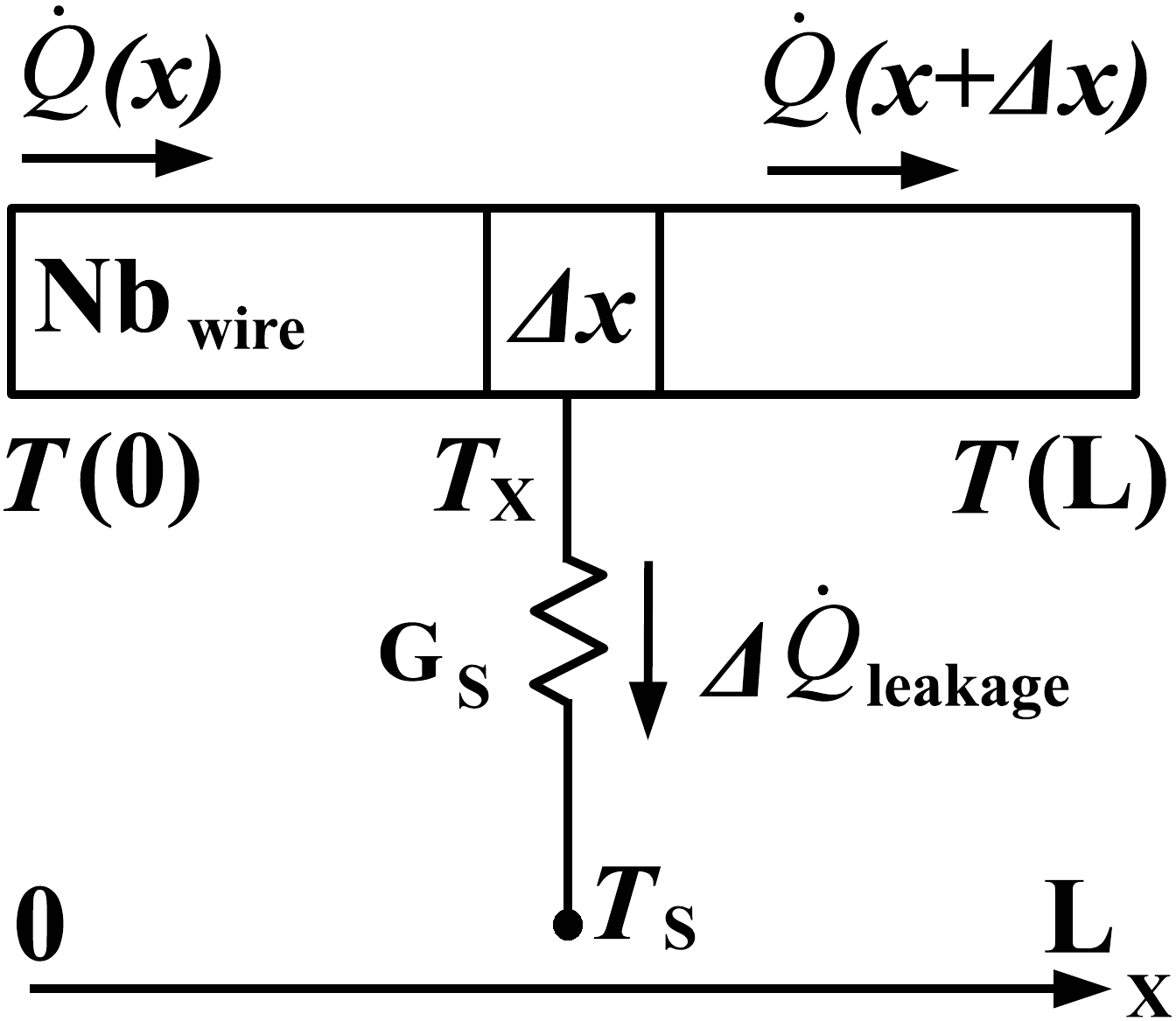}
\caption{Continuous model.}
\label{fig:10}
\end{figure}

Here, we calculate the heat flow in the structure by applying a continuous model shown in Fig.~\ref{fig:10}. We write the equation for the heat currents for small temperature differences as
\begin{equation}
\dot{Q}(x) = \dot{Q}(x + \Delta x) + \dot{Q}_{leakage}.
\label{Eq.10}
\end{equation}
One has $\frac{d\dot{Q}(x)}{dx} = -\frac{\Delta \dot{Q}_{leakage}}{\Delta x}$, where
\begin{equation}
\begin{split}
& \frac{d\dot{Q}(x)}{dx} = -G_{Nb}L\frac{d^{2}T(x)}{dx^{2}} & \\
& \frac{\Delta \dot{Q}_{leakage}}{\Delta x} = \frac{G_{S}}{L}(T(x) - T_{S}). &
\label{Eq.11}
\end{split}
\end{equation}
Using these definitions, one can obtain a second-order inhomogeneous differential equation  
\begin{equation}
\frac{d^{2}T(x)}{dx^{2}} - \frac{\rho}{L^{2}}T(x) = -\frac{\rho}{L^{2}}T_{S},
\label{Eq.12}
\end{equation}
with the boundary conditions $T(x) \big|_{0} = T_{L}$ and $T(x) \big|_{L} = T_{R}$. The temperature profile along the wire can be solved as $T(x) = T_{S} + (T_{L} - T_{S})\csch{\big(\sqrt{\rho}\big)}\sinh{\big(\frac{(L-x)\sqrt{\rho}}{L}\big)} + (T_{R} - T_{S})\csch{\big(\sqrt{\rho}\big)}\sinh{\big(\frac{x \sqrt{\rho}}{L}\big)}$.
In addition, the heat flows at the boundaries are
\begin{equation}
\begin{split}
& \dot{Q}_{wire}^{L} = G_{Nb}\sqrt{\rho}\big[(T_{L} - T_{S})\coth{(\sqrt{\rho})} - (T_{R} - T_{S})\csch{(\sqrt{\rho})}\big] & \\
& \dot{Q}_{wire}^{R} = G_{Nb}\sqrt{\rho}\big[(T_{L} - T_{S})\csch{(\sqrt{\rho})} - (T_{R} - T_{S})\coth{(\sqrt{\rho})}\big]. &
\label{Eq.13}
\end{split}
\end{equation}
Thus, the measured thermal conductance and the heat leak to the substrate are
\begin{equation}
\begin{split}
& G_{meas}\approx \sqrt{\rho}\csch{(\sqrt{\rho})}G_{Nb} & \\
& G_{leakage} \approx \sqrt{\rho}\tanh{(\sqrt{\rho}/2)}G_{Nb}. &
\label{Eq.14}
\end{split}
\end{equation}

\subsection*{I. Correction to the electron-phonon channel calibration due to additional heat leak to the substrate from the wire}

\begin{figure}[h!t]
\centering
\includegraphics[width=0.45\textwidth]{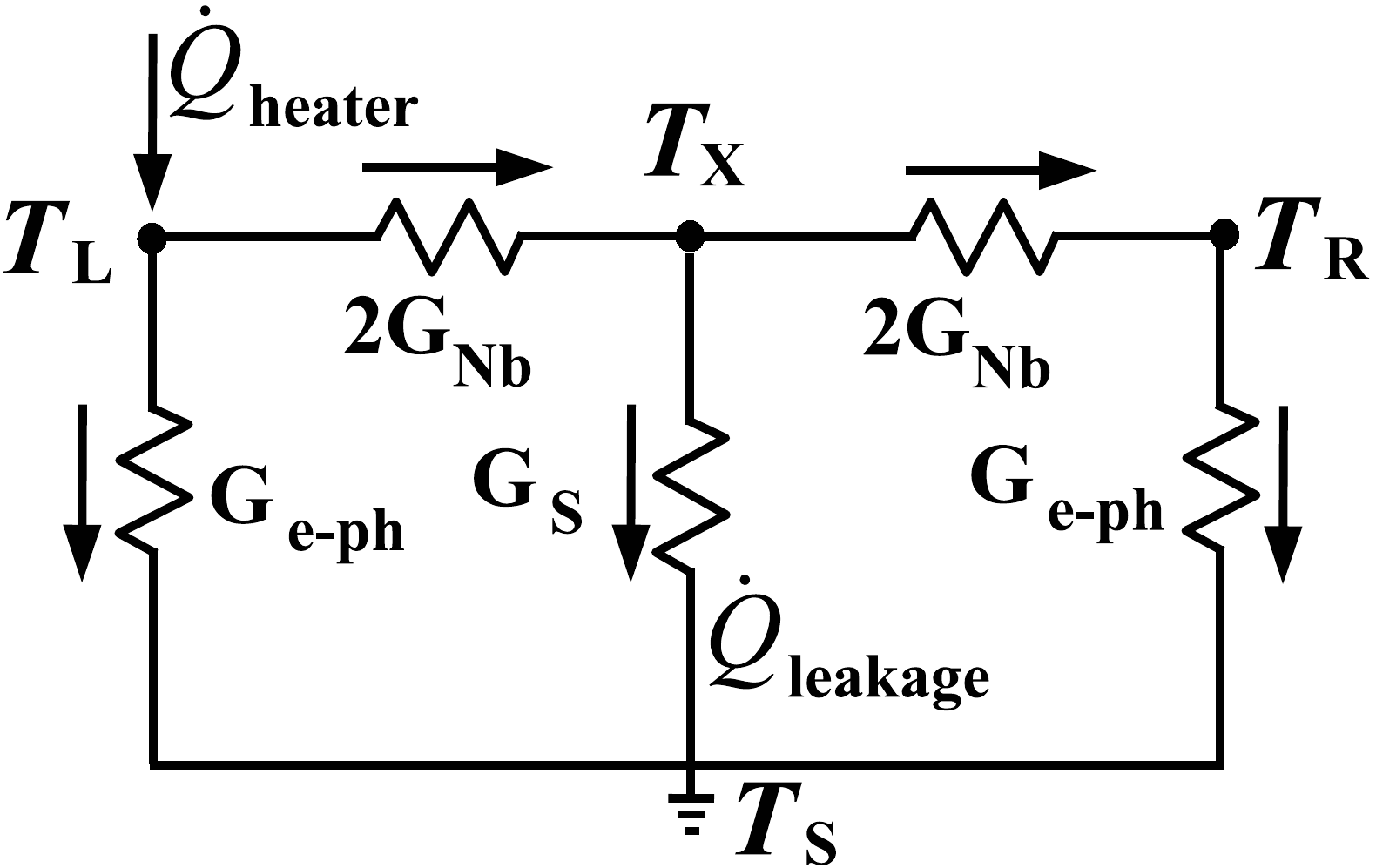}
\caption{Full model.}
\label{fig:11}
\end{figure}

Here, we consider a linearized model for the experimental setting where the left island is heated and the heat flows to the substrate via the $e-ph$ channel of the left Cu island, and to the Nb wire. Subsequently, heat is distributed from the wire to the substrate via heat leak $\dot{Q}_{leakage}$ and via the $e-ph$ channel of the right Cu island. A schematic of the considered heat flows is shown in Fig.~\ref{fig:11}.
A system of linear equations for the heat flow in the present structure reads
\begin{equation}
\begin{split}
& \dot{Q}_{heater} = G_{e-ph}(T_{L} - T_{S}) +2G_{Nb}(T_{L} - T_{x}) & \\
& 2G_{Nb}(T_{R} - T_{x}) + G_{e-ph}(T_{R} - T_{S}) = 0 & \\
& 2G_{Nb}(T_{L} - T_{x})+2G_{Nb}(T_{R} - T_{x}) - G_{S}(T_{x} - T_{S}) = 0 & \\
& \dot{Q}_{leakage} = G_{S}(T_{x} - T_{S}). &
\label{Eq.15}
\end{split}
\end{equation}
In order to evaluate the heat flow through the wire, we calculate a temperature profile as
\begin{equation}
\begin{split}
& T_{L} = \frac{\dot{Q}_{heater}+\frac{4\dot{Q}_{heater}}{4\gamma + \rho\gamma + 2\rho}}{G_{Nb}(2 + \gamma)} + T_{S} & \\
& T_{x} = \frac{2\dot{Q}_{heater}}{G_{Nb}(4\gamma + \rho\gamma + 2\rho)} + T_{S} & \\
& T_{R} = \frac{4\dot{Q}_{heater}}{G_{Nb}(\gamma + 2)(2\rho + \gamma(\rho + 4))} + T_{S}, &
\label{Eq.16}
\end{split}
\end{equation}
where $\rho = G_{S}/G_{Nb}$ and $\gamma = G_{e-ph}/G_{Nb}$.
Taking into account the heat leak to the substrate from the wire, we get the effective experimentally observed electron-phonon conductance as $G_{e-ph}^{eff} = \dot{Q}_{heater}/(T_{L} + T_{R} - 2T_{S})$, where
\begin{equation}
G_{e-ph}^{eff} = \frac{G_{Nb}(4\gamma + \rho\gamma + 2\rho)}{4 + \rho}.
\label{Eq.17}
\end{equation}
Note that $G_{e-ph}^{eff} = G_{e-ph}$ in case of absence of leakage, $G_{S} =~0$. Now, mimicking the experiment, we calculate the heat reaching the right island according to the effective $e-ph$ calibration as $\dot{Q}_{wire}^{R, eff} = G_{e-ph}^{eff}T_{R}$. At the same time, the temperature difference across the wire is $\Delta T = T_{L} - T_{R}$. Finally, the thermal conductance of the wire calculated using the effective $e-ph$ calibration is $G_{Nb}^{eff} = \dot{Q}_{wire}^{R, eff}/ \Delta T$, for which we obtain 
\begin{equation}
G_{Nb}^{eff} = \frac{4G_{Nb}}{4 + \rho}.
\label{Eq.18}
\end{equation}
Here, we obtain the same result as for the three-resistor model in Eq.~(\ref{Eq.9}), and $G_{Nb}^{eff} = G_{Nb}$, if $G_{S} = 0$.

\section*{Acknowledgements}

We acknowledge the availability of the facilities and technical support by Otaniemi research infrastructure for Micro and  Nanotechnologies (OMN). We acknowledge financial support from the European Community FP7 Marie Curie Initial Training Networks Action (ITN) Q-NET 264034 and INFERNOS grant (Project No. 308850), and the Academy of Finland grants (Projects No. 250280, No. 272218, No. 286098 and No. 275167). We thank Y. Galperin, C. Enss and D. Basko for useful discussions, and M. Meschke for technical assistance.

\section*{Author contributions statement}

A. V. F. fabricated Nb samples, conducted the experiments and wrote the manuscript, A. V. F. and O.-P. S. analyzed the data from measurements of Nb wires, J. T. P. provided the data from measurements of Al wires. All authors participated in designing the experiment and the development of thermal model, discussion of the results and reviewed the manuscript. 

\section*{Additional information}

\textbf{The authors declare no competing financial interests.}

\end{document}